\documentclass{PoS}

\title{On the Extraction of the Strong Coupling from Hadronic Tau Decay}

\ShortTitle{On the Extraction of the Strong Coupling from Hadronic Tau Decay}

\author{Diogo Boito\\
      Universitat Auton\`oma de Barcelona, Barcelona, Spain\\
}
 
\vspace*{-1pt}
\author{Oscar Cat\`a\\
      Universitat de Val\`encia and IFIC-CSIC, Val\`encia, Spain\\
      Arnold Sommerfeld Center for Theoretical Physics, LMU M\"unchen, Germany\\
}
 
\vspace*{-1pt}
\author{Maarten Golterman\\
      San Francisco State University, San Francisco, USA\\
}
 
\vspace*{-1pt}
\author{Matthias Jamin\\
      ICREA and IFAE, Universitat Auton\`oma de Barcelona, Barcelona, Spain\\
}
 
\vspace*{-1pt}
\author{Kim Maltman\\
      York University, Toronto, Canada\\
      CSSM, University of Adelaide, Adelaide, Australia\\
}
 
\vspace*{-1pt}
\author{\speaker{James Osborne}\\
      San Francisco State University, San Francisco, USA\\
}
 
\vspace*{-1pt}
\author{Santiago Peris\\
      San Francisco State University, San Francisco, USA\\
      Universitat Auton\`oma de Barcelona, Barcelona, Spain\\
}

\abstract{The finite energy sum rule extraction of the strong coupling $\alpha_s$ from hadronic $\tau$ decay data provides one of its most precise experimental determinations. As precision improves, small non-perturbative effects become increasingly relevant to both the central value and error. Here we present a new framework for this extraction employing a physically motivated model to accommodate violations of quark-hadron duality and enforcing a consistent treatment of higher-dimension operator product expansion contributions. Using 1998 OPAL data for the non-strange vector and axial-vector spectral functions, we find the $n_f=3$ values for $\alpha_s(m_\tau^2)$ of $0.307(19)$ for fixed-order perturbation theory and $0.322(26)$ for contour-improved perturbation theory, corresponding to $n_f=5$ values for $\alpha_s(M_Z^2)$ of $0.1169(25)$ and $0.1187(32)$, respectively.}

\FullConference{XXIX International Symposium on Lattice Field Theory \\
		 July 10--16 2011\\
		 Squaw Valley, Lake Tahoe, California}

\begin{document}

\section{Introduction}
\label{sec:introduction}

The world average of $\alpha_s$ is currently determined primarily by high-precision determinations on the lattice. However, in the past few years there has been a renewed interest in the precision determination of $\alpha_s$ from non-strange hadronic $\tau$ decays. One reason for this is the recent calculation of the $\mathcal{O}(\alpha_s^4)$ coefficient in the perturbative contribution to the Adler function \cite{Baikov}. The improved precision of the perturbative contribution emphasizes the importance of bringing other systematic effects, previously treated with varying degrees of completeness, under control. In fact, a number of competing analysis methods now obtain results which are not fully consistent with one another, even when applied to the same data.

There are several theoretical issues related to this discrepancy between different determinations. There is a question of which resummation scheme, contour-improved perturbation theory (CIPT) or fixed-order perturbation theory (FOPT), is best used in the analysis. This issue has been the focus of many recent analyses, and we have chosen to present the results of each scheme separately. The present analysis focuses on theoretical questions related to the non-perturbative contributions which, due to the relatively low value of the $\tau$ mass, are not entirely negligible. We present a new framework that both maintains consistency between the dimension at which the operator product expansion (OPE) is truncated and the choice of sum rule weights, and provides a quantitative description of duality violations (DVs). A complete account of the work presented here can be found in Ref. \cite{Boito}.

\section{Theory}
\label{sec:theory}

The analysis of hadronic $\tau$ decay begins with the ratio
\begin{equation}
R_\tau \equiv \frac{\Gamma \left [ \tau^- \rightarrow \nu_\tau \; \rm{hadrons} \right ]}{\Gamma \left [ \tau^- \rightarrow \nu_\tau \, e^- \, \overline{\nu}_e \right ]} \; .
\label{eq:ratio}
\end{equation}
Experimentally, it is possible to decompose the non-strange contribution to this branching ratio into vector (V) and axial-vector (A) components. Given the currents $J_V^\mu = \overline{u}(x) \gamma^\mu d(x)$ and $J_A^\mu = \overline{u}(x) \gamma^\mu \gamma^5 d(x)$, the two-point current correlation functions are defined by
\begin{eqnarray}
\Pi^{\mu \nu}_{V,A} &\equiv& i \int d^4x \; e^{i q \cdot x} \langle0| T \{ J^\mu_{V,A}(x) J^{\dagger \, \nu}_{V,A}(0) |0\rangle \\
&=& \left ( q^\mu q^\nu - q^2 g^{\mu \nu} \right ) \Pi_{V,A}^{(1)}(q^2) + q^\mu q^\nu \Pi_{V,A}^{(0)}(q^2) \; ,
\label{eq:correlator}
\end{eqnarray}
where the superscripts $(1,0)$ label spin. Defining the spectral functions $\rho_{V,A}^{(J)}(s) \equiv (1/\pi) \rm{Im} \; \Pi_{V,A}^{(J)}(s)$ with $s=q^2$ the invariant squared-mass of the hadronic system, the non-strange contributions to Eq.~(\ref{eq:ratio}) can also be expressed as the $s_0=m_\tau^2$ version of the weighted integral \cite{BNP}
\begin{equation}
R_{V,A;ud}(s_0) = S_{EW} |V_{ud}|^2 \frac{12 \pi^2}{s_0} \int_0^{s_0} ds \, \left ( 1 - \frac{s}{s_0} \right )^2 \left [ \left ( 1 + 2 \frac{s}{s_0} \right ) \rho_{V,A}^{(1+0)}(s) - 2 \left ( \frac{s}{s_0} \right ) \rho_{V,A}^{(0)}(s) \right ] \; .
\label{eq:nsratio}
\end{equation}
Since the $J=0$ contributions are $\mathcal{O}[(m_u \pm m_d)^2]$ suppressed and thus numerically negligible apart from the $\pi$ contribution to $\rho^{(0)}_A(s)$, the differential versions or Eq.~(\ref{eq:nsratio}) provide experimental determinations of $\rho^{(1+0)}_{V,A}(s)$.

\begin{figure}
\begin{minipage}{0.45\textwidth}
\begin{centering}
\includegraphics[width=0.7\textwidth]{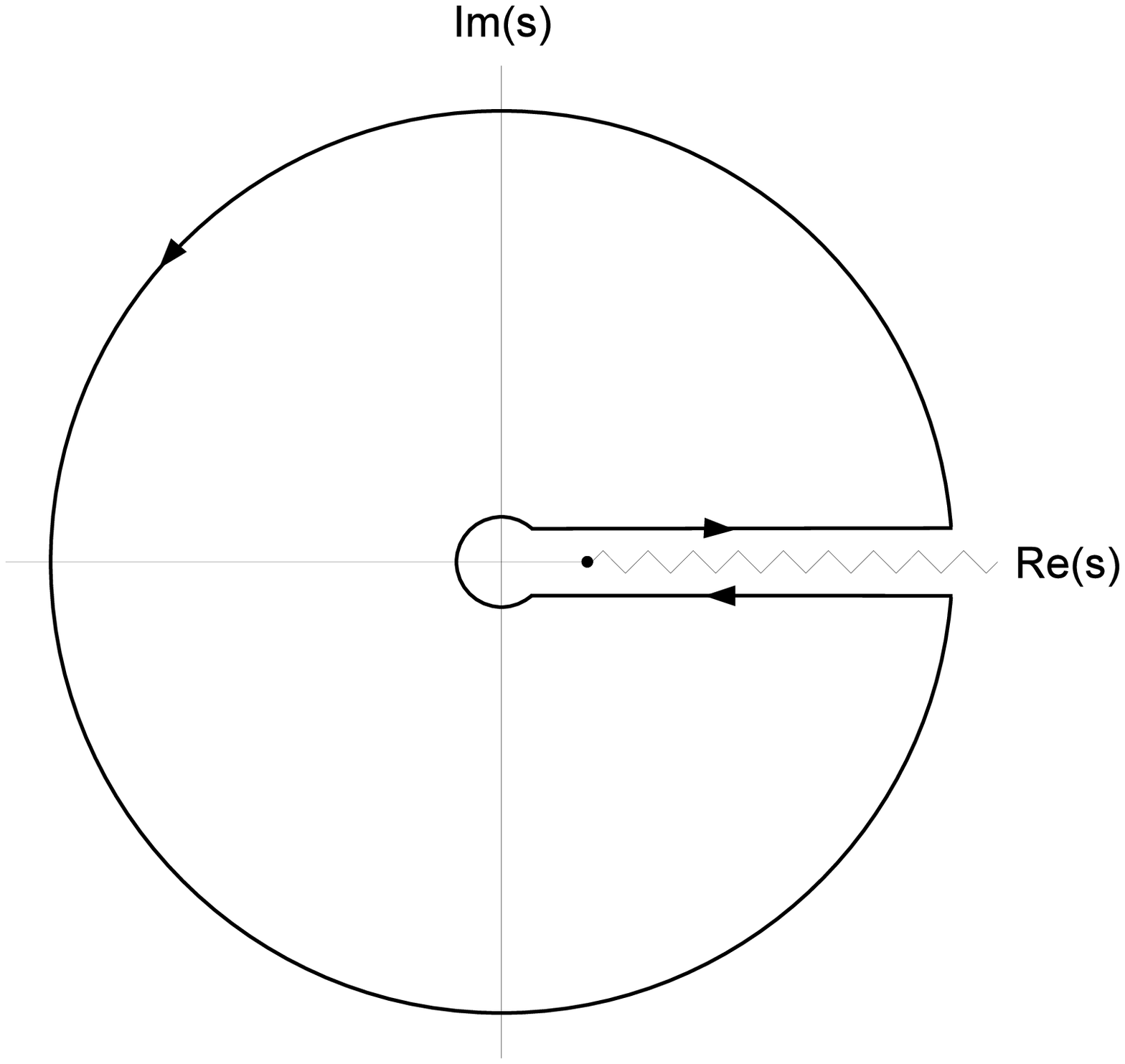}
\caption{\emph{FESR contour. The discontinuity across the cut is $2i \rm{Im} \Pi(s)$.}}
\label{fig:contour}
\end{centering}
\end{minipage}
\hspace*{0.082\textwidth}
\begin{minipage}{0.45\textwidth}
\begin{centering}
\includegraphics[width=0.7\textwidth]{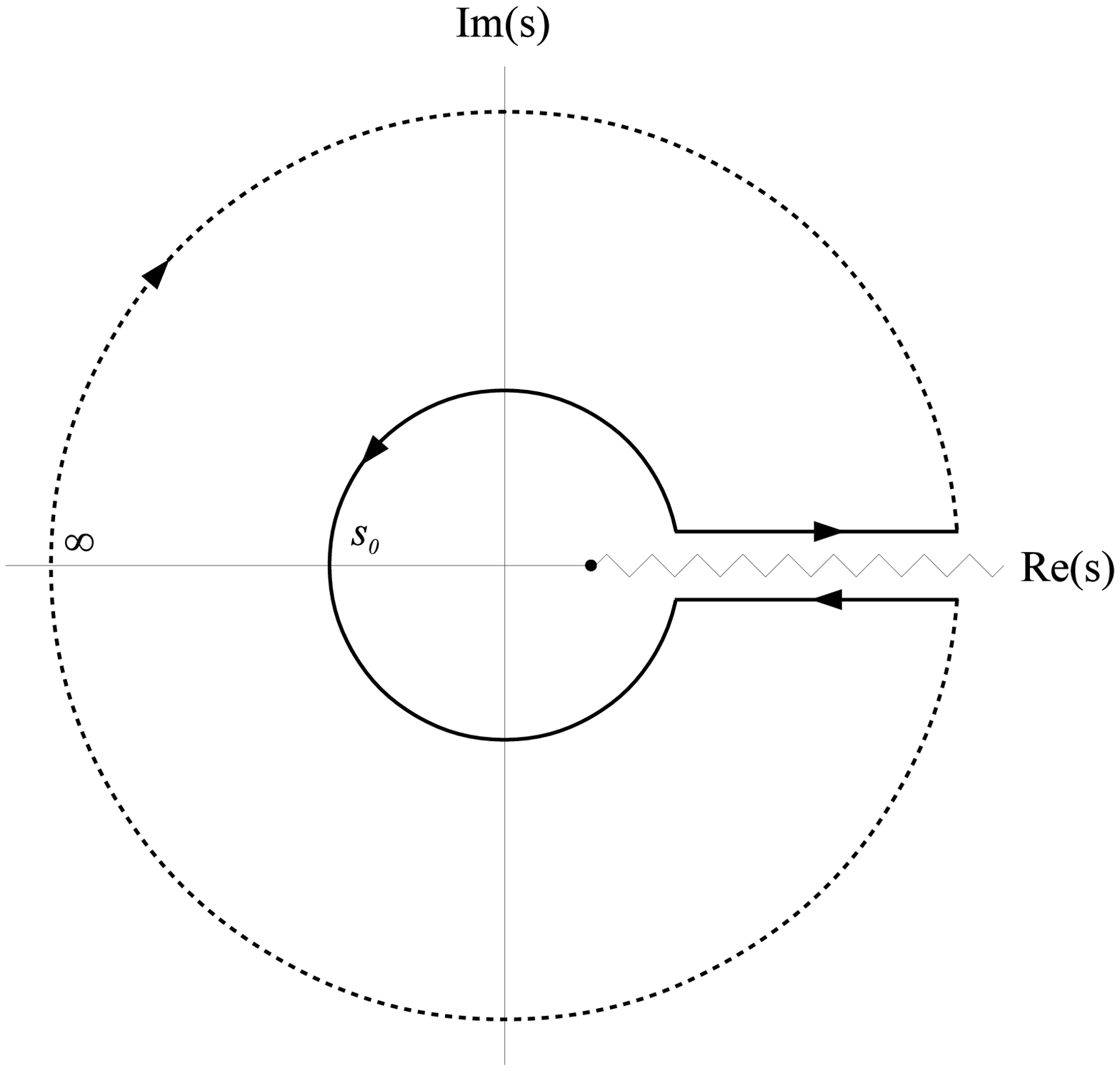}
\caption{\emph{Contour relating DVs to the spectral functions between $s_{min} \le s_0 \le \infty$.}}
\label{fig:contour2}
\end{centering}
\end{minipage}
\end{figure}

Cauchy's theorem, applied to the contour in Fig.~\ref{fig:contour}, implies that, for any $s_0$ and any analytic weight $w(s)$, $\Pi^{(1+0)}_{V,A}(s)$ satisfies the finite energy sum rule (FESR) relation \cite{Shankar}
\begin{equation}
\int_0^{s_0} ds \, w(s) \rho^{(1+0)}_{V,A}(s) = - \frac{1}{2 \pi i} \oint_{|s|=s_0} ds \, w(s) \Pi^{(1+0)}_{V,A}(s) \, .
\label{eq:fesr}
\end{equation}
The LHS of Eq.~(\ref{eq:fesr}) can be determined experimentally for $s_0 \le m_\tau^2$. This then allows QCD parameters such as $\alpha_s$ to be related to experimental data, provided the OPE representation
\begin{equation}
\Pi^{OPE}_{V,A}(s) = \Pi^{PT}_{V,A}(s) + \frac{C_4}{s^2} - \frac{C_6}{s^3} + \frac{C_8}{s^4} - \dots
\label{eq:ope}
\end{equation}
can be reliably employed on the RHS.\footnotemark This requires not only that $s_0 \gg \Lambda_{QCD}^2$, but also caution in treating contributions in the vicinity of the positive real $s$-axis, where the OPE is expected to breakdown for values of $s$ below $m_\tau^2$.

\footnotetext{Here we have ignored numerically negligible $\mathcal{O}(m_{u,d}^2)$ dimension 2 contributions. We will also ignore the logarithmic scale dependence of the $C_i$. For a more detailed discussion, see Ref. \cite{Boito}.}

Defining the correction to the OPE caused by DVs as $\Delta_{V,A}(s) = \Pi_{V,A}^{(1+0)}(s) - \Pi_{V,A}^{OPE}(s)$, we may relate the experimental spectral functions to the OPE by
\begin{eqnarray}
\int_0^{s_0} ds \, w(s) \rho^{(1+0)}_{V,A}(s) &=& -\frac{1}{2 \pi i} \oint_{|s|=s_0} ds \, w(s) \left [ \Pi_{V,A}^{OPE}(s) + \Delta_{V,A}(s) \right ] \\
&=& -\frac{1}{2 \pi i} \oint_{|s|=s_0} ds \, w(s) \Pi_{V,A}^{OPE}(s) - \int_{s_0}^\infty ds \, w(s) \rho_{V,A}^{DV}(s) \; ,
\label{eq:dv_relation}
\end{eqnarray}
where $\rho_{V,A}^{DV} \equiv (1/\pi) \Delta_{V,A}(s)$. The second line of Eq.~(\ref{eq:dv_relation}) comes from an application of Cauchy's theorem using the contour in Fig.~\ref{fig:contour2}, provided $\rho^{DV}_{V,A}(s)$, as expected, vanish fast enough as $s \rightarrow \infty$ \cite{CGP1}. To suppress DVs, previous analyses have typically employed only doubly-pinched weights, and often restricted their attention to $s_0=m_\tau^2$. With these precautions, all previous analyses have taken $\rho_{V,A}^{DV}(s) $ to be negligible.

\section{Strategy}
\label{sec:strategy}

In actual analyses, it is necessary to truncate the non-perturbative OPE series, both for practical reasons and because the series is likely asymptotic. Through the residue theorem, a term of order $s^n$ in the polynomial $w(s)$ will pick out the term in the OPE of order $1/s^{n+1}$. Consistency then requires that, if the OPE is truncated at $\mathcal{O}(1/s^{n+1})$, the maximum degree of the polynomial weights employed should be $n$ \cite{Maltman}. OPE contributions for weights of degree $n$ will in general depend on the $n+1$ OPE parameters $\alpha_s$, $C_4$, $C_6$, $\dots$, $C_{2n+2}$. As the number of such linearly independent weights is also $n+1$, one can not perform true fits using FESRs involving only a single $s_0$ value while maintaining such consistency. Our analysis therefore involves the range $s_{min} \le s_0 \le m_\tau^2$, with $s_{min}$ chosen such that theory gives a good description of the data above it.

We will also consider the presence of DVs in this analysis. While quite generally one would like to understand the effect of DVs, the necessity of working with $s_0 < m_\tau^2$ makes their inclusion more relevant since at lower $s_0$ values the residual DV effects are expected to be larger. Since no systematic theory of DVs is currently available from QCD, we employ a physically motivated model to examine their contributions. This \emph{ansatz}, based on Regge theory, large-$N_c$, and analyticity \cite{Blok}, is given by \cite{CGP1, CGP2}
\begin{equation}
\rho_{V,A}^{DV}(s) = \kappa_{V,A} \, e^{-s \, \gamma_{\, V,A}} \sin \left ( \alpha_{V,A} + s \, \beta_{V,A} \right ) \, .
\label{eq:ansatz}
\end{equation}
Introducing the DV \emph{ansatz} above adds four new fit parameters per channel on top of the OPE parameters considered in earlier studies. The data needed to fit this expanded parameter set is generated by evaluating the LHS of Eq.~(\ref{eq:dv_relation}) for a range of $s_0$ with a set of weights $w(s)$.

Including the \emph{ansatz}, Eq.~(\ref{eq:ansatz}), in this analysis allows us to perform fits with polynomial weights $w(s)$ that are less than doubly-pinched. In the following section, we first present the results of fits to the unpinched weight $w(s)=1$. After demonstrating the ability of our model to describe unpinched sum rules, we then present results of fits involving multiple weights, both pinched and unpinched. By showing, in the latter fits, our model's ability to describe sum rules with a diverse variety of weights, we aim to establish that our model for DVs provides an accurate description of physics missing from the OPE representation of the correlators.

\section{Fits}
\label{sec:results}

Two experiments have made public their $\tau$ decay spectral functions, ALEPH \cite{ALEPH, ALEPH2} and OPAL \cite{OPAL}. The 2005 analysis of ALEPH is more recent and is based on more statistics, so it would be the natural choice to apply our fitting scheme. Unfortunately, the publicly available ALEPH data omits correlations due to unfolding. Since a reanalysis of this data by the ALEPH collaboration is ongoing, the present analysis will focus on the 1998 OPAL data.

\begin{figure}
\begin{tabular}{c c}
\includegraphics[width=0.47\textwidth]{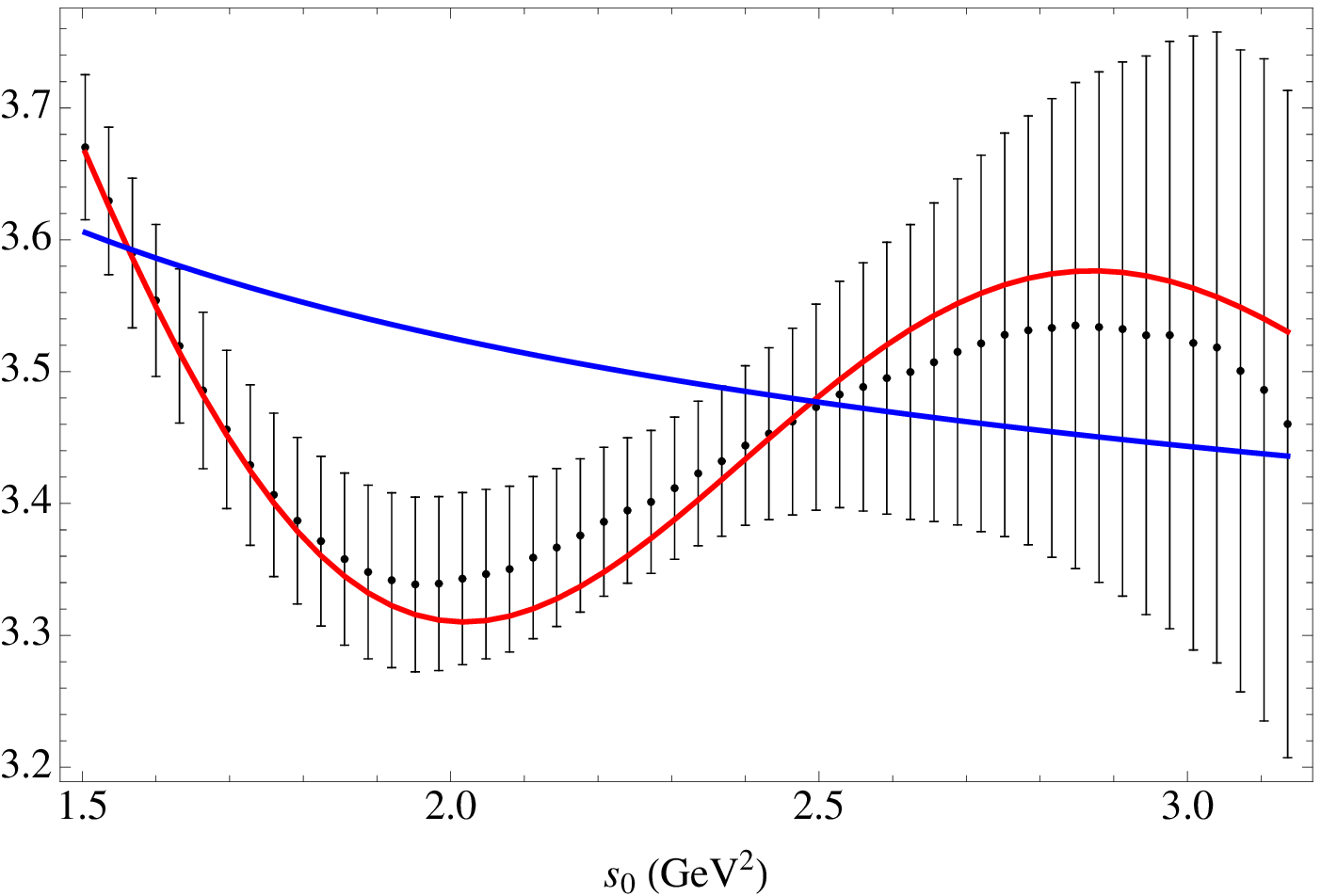}
&
\includegraphics[width=0.47\textwidth]{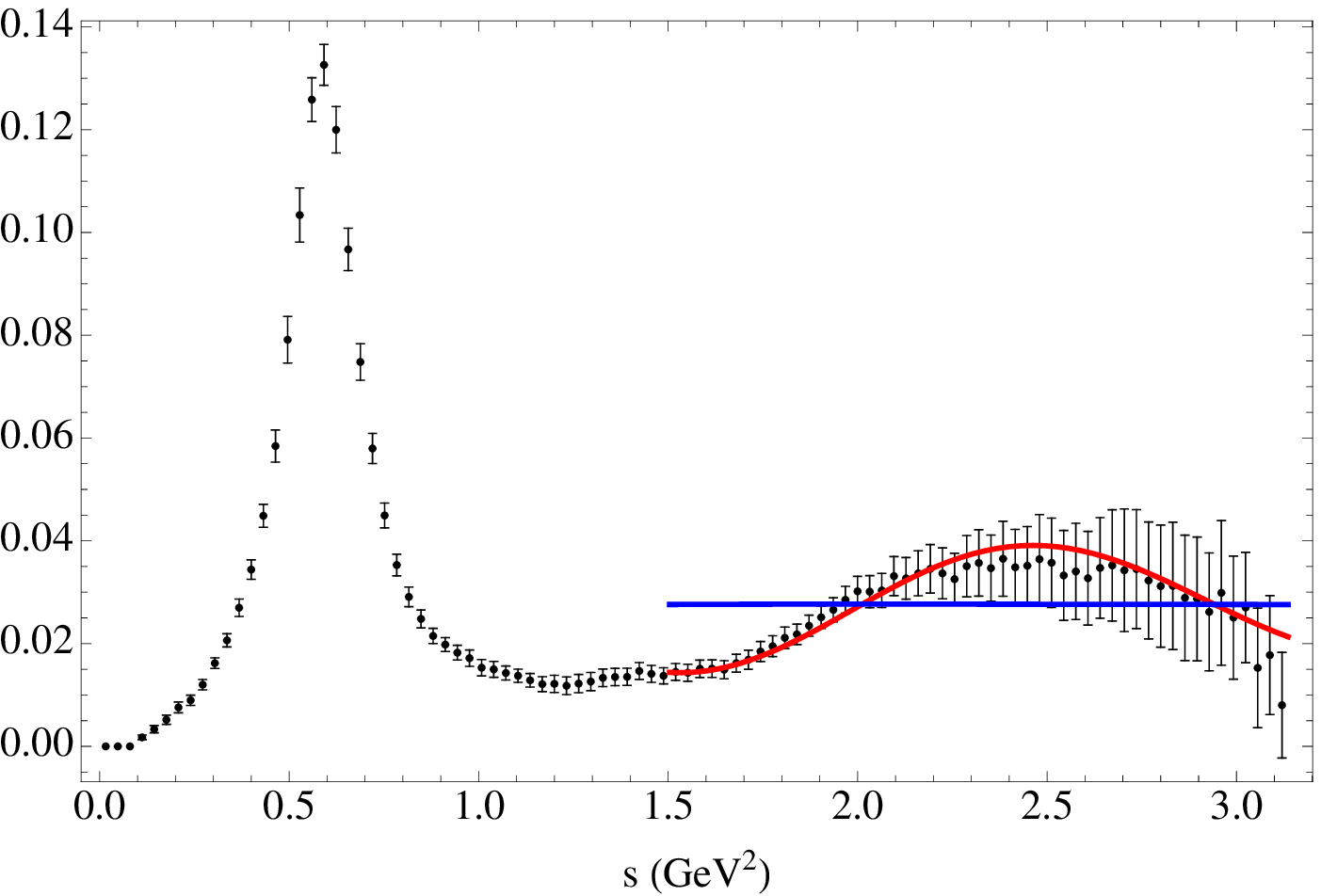}
\end{tabular}
\caption{\emph{The result of FOPT fits to V channel data with $w(s)=1$. Shown are, on the left, the OPE(+DV) and spectral integrals and, on the right, the the OPE(+DV) and OPAL experimental $\rho_V^{(1+0)}(s)$. Red curves give the full OPE+DV results, while blue curves show only the OPE contributions.}}
\label{fig:w1vector}
\end{figure}

We first present the results of standard $\chi^2$ fits to Eq.~(\ref{eq:dv_relation}) with data generated in the interval $1.5 \; \rm{GeV}^2 \le s_0 \le m_\tau^2$, where the lower bound is chosen based on fit quality and stability. Fig. \ref{fig:w1vector} shows the results of fits to OPAL V channel data using the weight $w(s)=1$, which results in values for the strong coupling of
\begin{eqnarray}
&\alpha_s(m_\tau^2)& = 0.307 \pm 0.019 \phantom{abcdef} \rm{(FOPT)} \; , \\
&\alpha_s(m_\tau^2)& = 0.322 \pm 0.026 \phantom{abcdef} \rm{(CIPT)} \; .
\label{eq:w1vector}
\end{eqnarray}
The figures demonstrate not only the expected presence of duality violations in the V channel but also the ability of the ansatz, Eq.~(\ref{eq:ansatz}), to accurately describe the data. Plots of CIPT results are nearly identical to the FOPT results shown here.

\begin{figure}
\begin{tabular}{c c}
\includegraphics[width=0.47\textwidth]{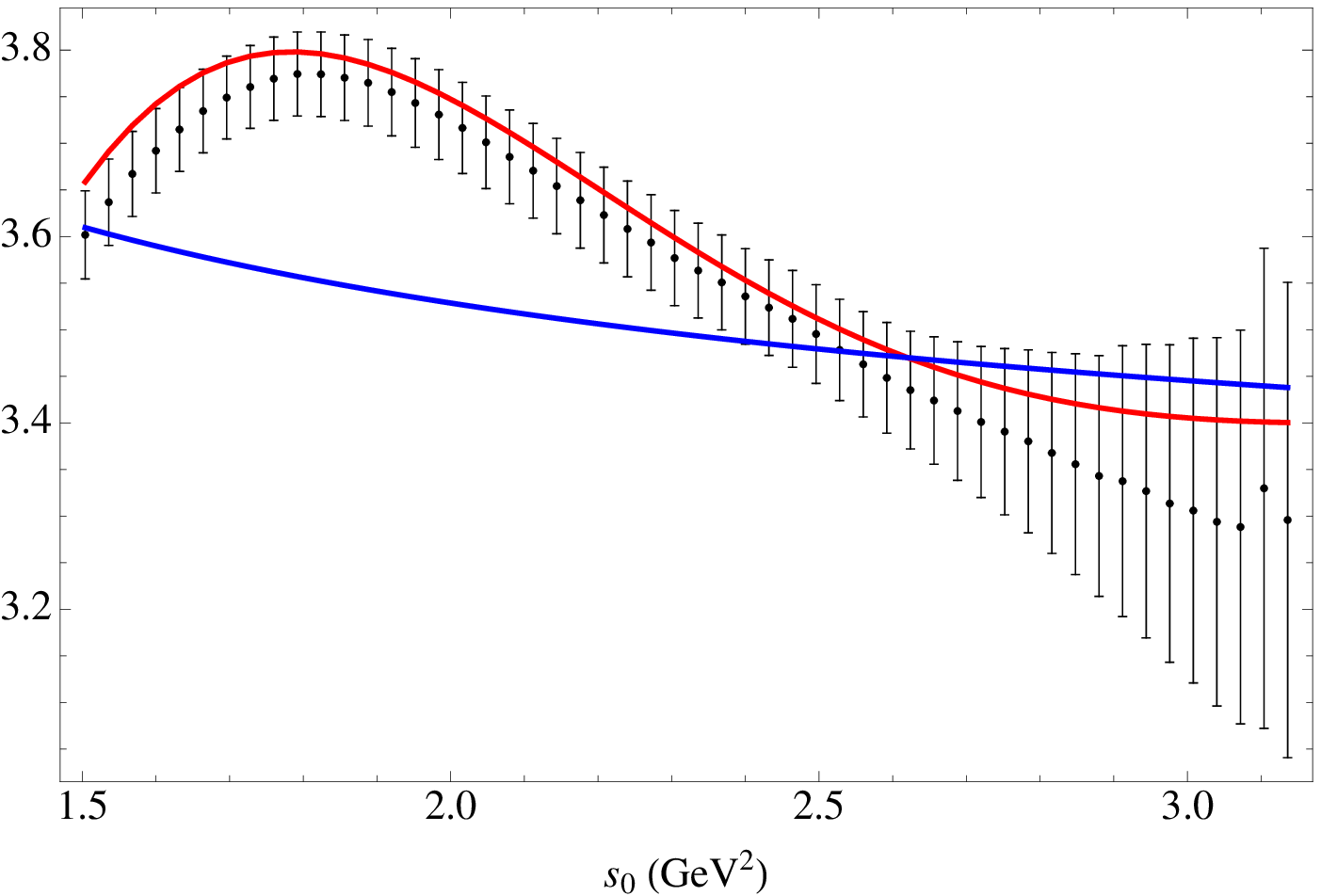}
&
\includegraphics[width=0.47\textwidth]{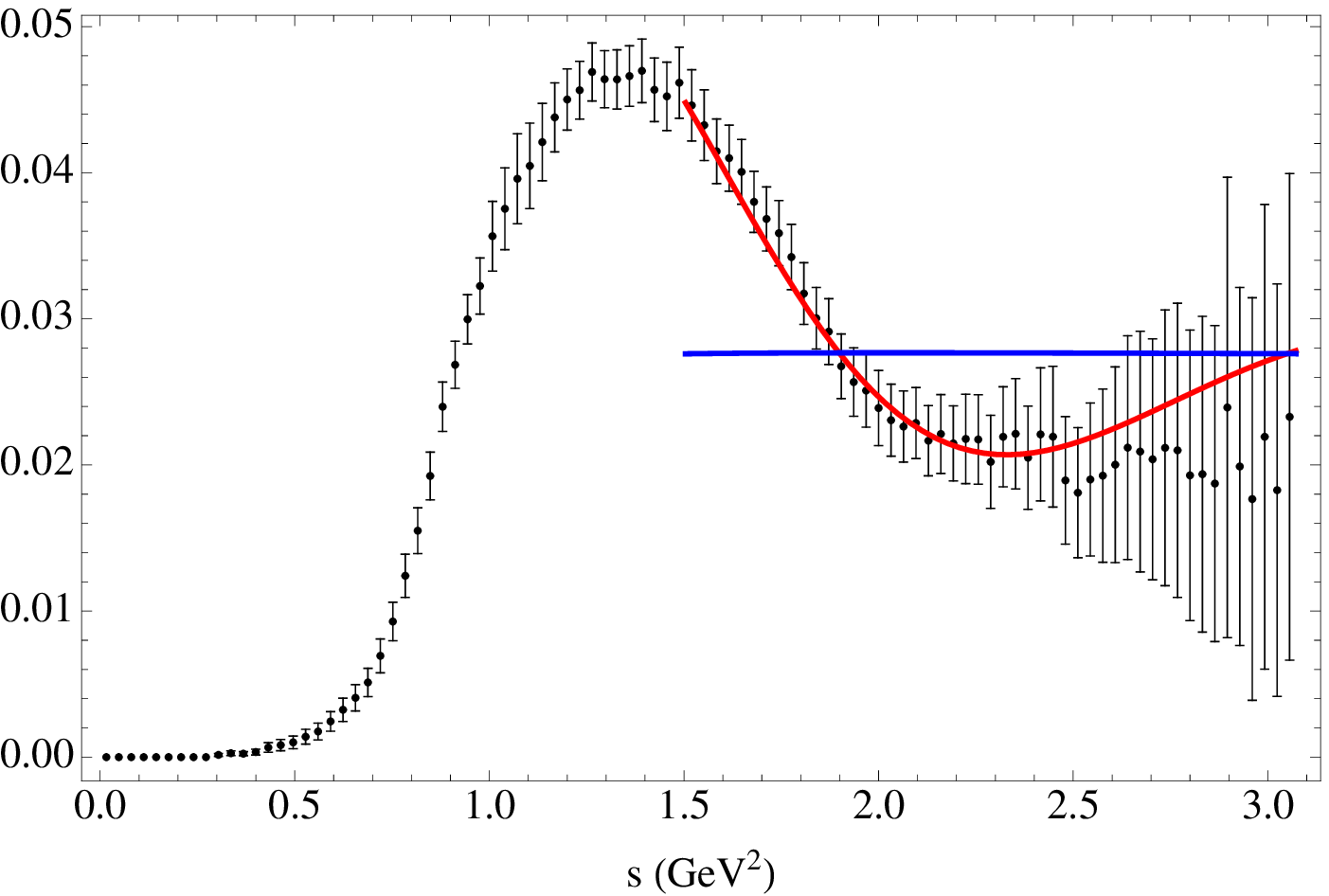}
\end{tabular}
\caption{\emph{The result of FOPT fits to V and A channel data with $w(s)=1$. Shown are, on the left, axial OPE(+DV) and spectral integrals and, on the right, the OPE(+DV) and OPAL experimental $\rho_A^{(1+0)}(s)$. Red curves give the full OPE+DV results, while blue curves show only the OPE contributions.}}
\label{fig:w1axial}
\end{figure}

A simultaneous fit to OPAL V and A channel data over the same $s_0$ interval using $w(s)=1$ yields values for $\alpha_s(m_\tau^2)$ of $0.308 (18)$ and $0.325 (25)$ for FOPT and CIPT, respectively. These values are completely consistent with the V-only results of Eq.~(\ref{eq:w1vector}). Fig. \ref{fig:w1axial} again demonstrates not only the expected presence of DVs in the A channel but also the ability of our model to accurately describe the data. Fits to the A channel alone, however, are less stable than the V-only case.

To examine the ability of the \emph{ansatz} to describe sum rules involving pinched weights we must refine our strategy. Fitting the parameters of Eq.~(\ref{eq:ansatz}) using only a single pinched weight leads to unstable results due to the pinched weight's suppression of DVs. To proceed, we fit to data generated by more than one weight, always including the unpinched weight $w(s)=1$ to fix the parameters of the DV model. The strong correlations between data generated using a range of $s_0$ and different $w(s)$, however, lead to correlation matrices with eigenvalues which are zero at machine precision, preventing a standard $\chi^2$ construction from being employed. To deal with this problem, we fit our set of OPE and DV parameters $\vec{p}$ by minimizing the alternate fit quality
\begin{equation}
Q^2 = \sum_w \sum_{s_0^i, s_0^j} \left ( I_{ex}^{(w)}(s_0^i) - I_{th}^{(w)}(s_0^i;\vec{p}) \right ) \left ( C^{(w)} \right )_{ij}^{-1} \left ( I_{ex}^{(w)}(s_0^j) - I_{th}^{(w)}(s_0^j;\vec{p}) \right ) \; ,
\label{eq:q2}
\end{equation}
where $I_{ex}^{(w)}$ is the weighted spectral integral, $I_{th}^{(w)}$ is the weighted theory integral which depends on the set of parameters $\vec{p}$, and $C^{(w)}$ is the full covariance matrix for the weight $w$.

This approach allows us to study the ability of the \emph{ansatz} to describe data simultaneously for a range of pinched and unpinched weights. $Q^2$ of course differs from the standard $\chi^2$ function, as the latter includes also correlations between spectral integrals involving different weights. Being forced to work with $Q^2$ means we lose the statistical interpretation of the distribution around the minimum of our fit quality present for the $\chi^2$ function. Care must then be taken to properly incorporate the effects of such ``cross-moment'' correlations on the errors and covariances of the fitted parameters. This is done here by performing a linear fluctuation analysis as detailed in Ref. \cite{Boito}.

\begin{figure}
\begin{tabular}{c c}
\includegraphics[width=0.47\textwidth]{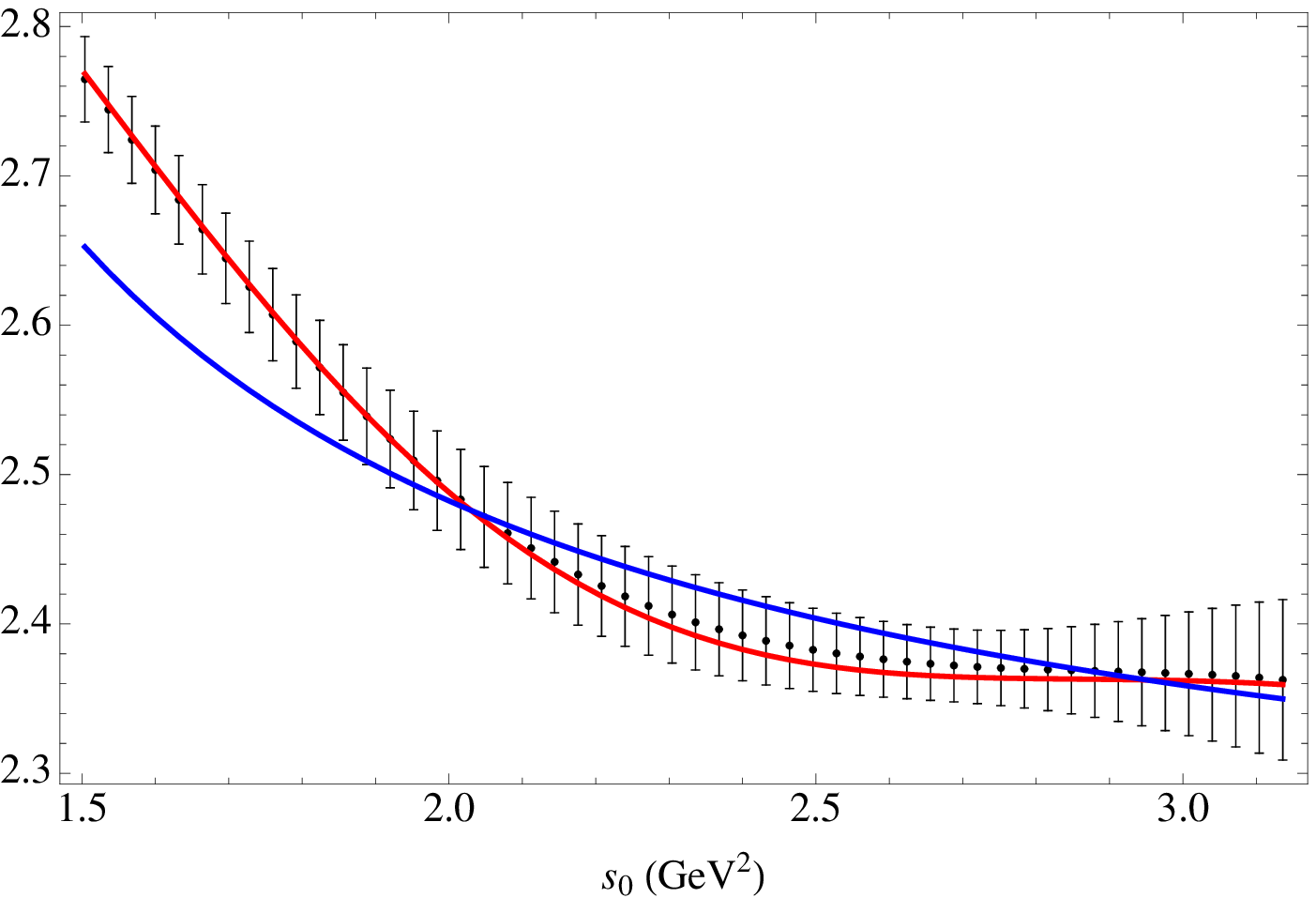}
&
\includegraphics[width=0.47\textwidth]{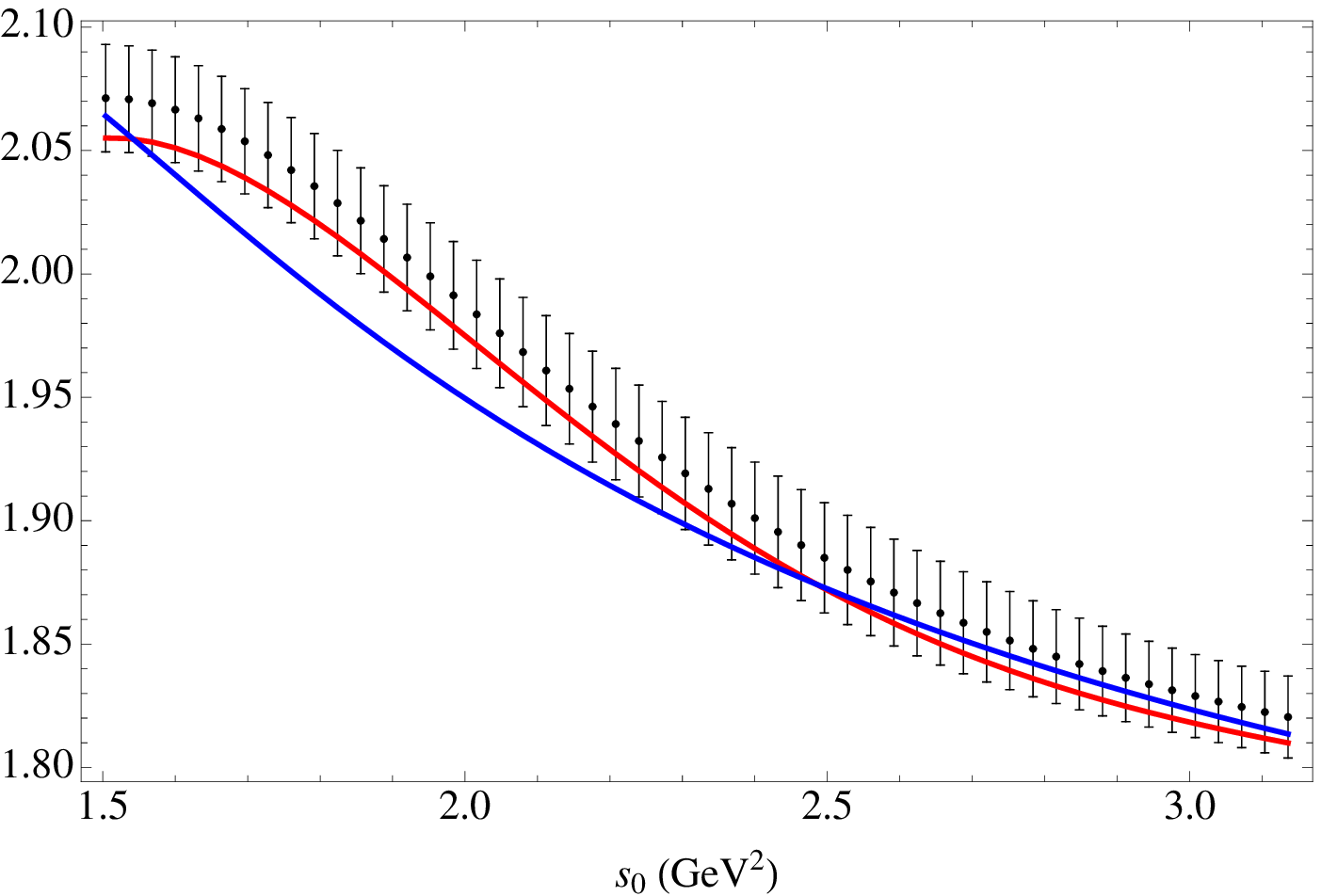}
\end{tabular}
\caption{\emph{The results of FOPT fits to V channel data with $w_0(s)=1$, $w_2(s)=1-(s/s_0)^2$, and $w_3(s)=(1-s/s_0)^2(1+2s/s_0)$. The left and right plots show the OPE(+DV) and spectral integrals for $w_2(s)$ and $w_3(s)$, respectively. Red curves give the full OPE+DV results, while blue curves show only the OPE contributions.}}
\label{fig:w123vector}
\end{figure}

Fitting the OPAL V channel data over the range $1.5 \; \rm{GeV}^2 \le s_0 \le m_\tau^2$ using the fit quality~(\ref{eq:q2}) and the set of weights $w_0(s) = 1$, $w_2(s) = 1 - (s/s_0)^2$, and $w_3(s) = (1 - s/s_0)^2(1+2s/s_0)$ yields values for $\alpha_s(m_\tau^2)$ of $0.304 (19)$ and $0.322 (31)$ for FOPT and CIPT, respectively. Fig. \ref{fig:w123vector} shows the results of this analysis for the $w_2$ and $w_3$ cases (the corresponding $w_0$ results are indistinguishable from those shown in Fig. \ref{fig:w1vector}). We find that our DV \emph{ansatz} is able to provide an accurate simultaneous description of the experimental spectral integrals for all of the weights employed, whether unpinched, singly-pinched, or doubly-pinched. Though not shown here explicitly, we find that an accurate simultaneous description is not possible using OPE contributions alone. The results for $\alpha_s$ from the combined $w_0$, $w_2$, and $w_3$ fit are clearly in excellent agreement with those from the single-weight, unpinched $w_0$ fit shown in Eq.~(\ref{eq:w1vector}). While we have chosen to present the result of a fit using the specific basis of weights described above, a variety of different weight sets have been examined with excellent consistency throughout.

\section{Conclusions}
\label{sec:conclusions}

We have presented an improved strategy for FESR analyses of hadronic $\tau$ decay data. First, the analysis enforces consistency between the degree of the weight employed and the dimension at which the OPE is truncated. Second, the analysis allows for a quantitative investigation of the impact of residual duality violations not present in previous analyses. This is accomplished by employing a model, Eq.~(\ref{eq:ansatz}), for the duality violation contributions to the vector and axial-vector channel spectral functions. By examining the stability of fits relative to both the range of energies used in the sum rules and the choice of weights, we conclude that our model is able to accurately describe the physics of duality violations in the vector and axial-vector channels.

Our most stable results come from fits to only the vector channel data. Regardless, all of the results obtained from combined vector and axial-vector fits are consistent with the vector channel only results. We take as our main result the standard $\chi^2$, $w(s)=1$, vector channel output of Eq.~(\ref{eq:w1vector}), $\alpha_s(m_\tau^2) = 0.307(19)$ for FOPT and $\alpha_s(m_\tau^2) = 0.322(26)$ for CIPT. For reference, the values of $\alpha_s(m_\tau^2)$ obtained by OPAL from the same data are $0.324(14)$ and $0.348(21)$ for FOPT and CIPT, respectively. Our results are seen to have central values shifted downward by slightly more than $1\\signa$, somewhat larger errors, and a somewhat smaller FOPT-CIPT difference. The larger errors are not unexpected, given the necessity of fitting four extra parameters per channel.

Running our values up to the $Z$ mass yields
\begin{eqnarray}
&\alpha_s(M_Z^2)& = 0.1169 \pm 0.0025 \phantom{abcdef} \mbox{($\overline{MS}$, $n_f=5$, FOPT)} \; , \\
&\alpha_s(M_Z^2)& = 0.1187 \pm 0.0032 \phantom{abcdef} \mbox{($\overline{MS}$, $n_f=5$, CIPT)} \; .
\end{eqnarray}
The slight asymmetry caused by scaling the errors has been averaged over.

\section{Acknowledgements}

We would like to thank Martin Beneke, Claude Bernard, Andreas H\"ocker, Manel Martinez, and Ramon Miquel for useful discussions.  We also would like to thank Sven Menke for significant help with understanding the OPAL spectral-function data. This work was supported in part by NSERC (Canada), the Ministerio de Educaci\'on (Spain), and the Dept. of Energy (USA).


\begin{thebibliography}{99}

 \bibitem{Baikov}
 P.~Baikov, K.~Chetyrkin, J.~K\"uhn,
 Phys. Rev. Lett. {\bf101}, 012002 (2008)
 [arXiv:0801.1821].
 
 \bibitem{Boito}
D.~Boito et al.
[arXiv:1110.1127v1 [hep-ph]].

\bibitem{BNP}
E.~Braaten, S.~Narison, A.~Pich,
Nucl. Phys. B {\bf 373} 581 (1992).

\bibitem{Shankar}
See, e.g., R.~Shankar,
Phys. Rev. D {\bf 15} 755 (1977).

\bibitem{LeDiberder}
F.~LeDiberder, A.~Pich,
Phys Lett. B {\bf 289} 165 (1992).

\bibitem{CGP1}
O.~Cat\`a, M.~Golterman, S.~Peris,
JHEP {\bf 0508}, 076 (2005)
[hep-ph/0506004].

\bibitem{Maltman}
K.~Maltman, T.~Yavin,
Phys. Rev. D {\bf 78} 094020 (2008)
[arXiv:0812.2285 [hep-ph]].

\bibitem{Blok}
B.~Blok, M.~Shifman, D.~Zhang,
Phys. Rev. D {\bf 57}, 2691 (1998)
[arXiv:hep-ph/9709333].

\bibitem{CGP2}
O.~Cat\`a, M.~Golterman, S.~Peris,
Phys. Rev. D {\bf 77}, 093006 (2008)
[arXiv:0803.0246 [hep-ph]].

\bibitem{ALEPH}
R.~Barate et al. [ALEPH Collaboration],
Eur. Phys. J. C {\bf 4}, 409 (1998).

\bibitem{ALEPH2}
S.~Schael et al. [ALEPH Collaboration],
Phys. Rept. {\bf 421}, 191 (2005)
[arXiv:hep-ex/0506072].

\bibitem{OPAL}
K.~Ackerstaff et al. [OPAL Collaboration],
Eur. Phys. J. C {\bf 7}, 571 (1999)
[arXiv:hep-ex/9808019].
 
\end{thebibliography}
\end{document}